\documentclass[%
 reprint,
superscriptaddress,
 amsmath,amssymb,
 aps,
 twocolumn,
]{revtex4}

\usepackage{graphicx}
\usepackage{dcolumn}
\usepackage{bm}
\usepackage{color}

\begin{document}

\preprint{APS/123-QED}

\title{Measuring surface tensions of soft solids with huge contact-angle hysteresis}


\author{Jin Young Kim}
\affiliation{Department of Materials, ETH Z\"{u}rich}%

\author{Stefanie Heyden} \affiliation{Department of Materials, ETH Z\"{u}rich}%

\author{Dominic Gerber} \affiliation{Department of Materials, ETH Z\"{u}rich}%

\author{Nicolas Bain} \affiliation{Department of Materials, ETH Z\"{u}rich}%
 
\author{Eric R. Dufresne}%
\affiliation{Department of Materials, ETH Z\"{u}rich}%

\author{Robert W. Style}
\affiliation{Department of Materials, ETH Z\"{u}rich}%
\email{robert.style@mat.ethz.ch}

\date{\today}

\begin{abstract}
The equilibrium contact angle  of a droplet resting on a solid substrate can reveal essential properties of the solid's surface.
However, when the motion of a droplet on a surface shows significant hysteresis,  it is generally accepted that the solid's equilibrium properties cannot be determined.
Here, we describe a method to measure surface tensions of soft solids with strong wetting hysteresis.
With independent knowledge of the surface tension of the wetting fluid and the linear-elastic response of the solid, the solid deformations under the contact line and the contact angle of a single droplet   together reveal  the difference in surface tension of the solid against the liquid and vapor phases.  
If the solid's elastic properties are unknown, then this surface tension difference can be determined from the change in substrate deformations with contact angle.
These results reveal an alternate equilibrium contact angle, equivalent to the classic form of Young-Dupr\'{e}, but with surface tensions in place of surface energies.
We motivate and apply  this approach with experiments on gelatin, a common hydrogel.

\end{abstract}

\maketitle


The most simple and widely used  technique to characterize a solid surface is the measurement of the contact angle of sessile droplets.
This is typically achieved by relating the equilibrium contact angle, $\theta_{eq}$, to the surface-energy difference across a contact line \emph{via} the law of Young-Dupr\'{e}:
\begin{equation}
    \gamma_{lv} \cos{\theta_{eq}} = \gamma_{sv}-\gamma_{sl}.
    \label{eq:young}
\end{equation}
Here, $\gamma$ represents surface energy, and subscripts $s,v,l$ correspond to solid, vapor and liquid respectively.
Absolute values of surface energies can be estimated by using information from wetting experiments with multiple different liquids (\emph{e.g.} \cite{owen69}).
On many surfaces, however, it is  impossible to accurately measure equilibrium contact angles:
the contact angle exhibits hysteresis due, for example, to surface roughness, pinning, or surface adaption \cite{dege04}.
Then, only advancing and receding contact angles can be measured reliably \cite{huhtamaki2018surface}.
Hysteresis-free surfaces  are rare in the lab, and even rarer in   real-world applications.
Thus,  quantitative approaches are needed to characterize the surface properties of hysteretic surfaces.

This overarching challenge is reflected in the more specific case of wetting on soft surfaces.
In such systems,  contact lines drive significant deformation of solid substrates, typically a swollen polymer network.
Revelevant materials include  hydrogels, silicones, organogels, ionogels, rubbers, and thermoplastic elastomers.
In general, such surfaces are hysteretic.
However, with few exceptions \cite{extr96,kaji13}, the study of soft wetting has focused on silicone gels and elastomers, using immiscible liquids such as water or glycerol (\emph{e.g.} \cite{carr96,peri08,jeri11,styl13,van2020}).
These show 
little-to-no hysteresis  \cite{xu17}, and thus avoid the problems inherent in working with `non-ideal' surfaces.

Hydrogels are a very important class of soft solids.
They can take a variety of forms, and are widespread due to their central role in biological tissue.
Their wetting properties are rich but difficult to measure quantitatively.
For example, it may seem natural that water should spread on hydrogels, since they are  made mostly of the same liquid. 
However, water droplets often take finite contact angles on hydrogel surfaces \cite{boul16}, and can exhibit large contact-angle hysteresis \cite{holl76}.
Hydrogel surfaces can also adapt to their environment, redistributing the network \cite{yao16,meie19} and reorienting its subunits
\cite{yasu81}, which give rise to dynamic surface properties.
Physically cross-linked hydrogels, typically  found in, or derived from living systems, can undergo time-dependent changes in the network structure \cite{bai19}.
Finally, hydrogels can  dry out, swell, and even dissolve during wetting
\cite{monteux2009role}.

Here,  we introduce  a method to measure surface tension differences across contact lines of hysteretic soft materials, based on the measurement of the microscopic deformations of the gel upon wetting.
Essentially, this is achieved by measuring $\theta_0$, the angle at which the surface tension forces acting at the contact line are in horizontal force balance.
This approach requires knowledge of the liquid-vapor surface tension of the wetting liquid, but the elastic constants of the gel can be unknown.
We demonstrate this approach with gelatin, a ubiquitous biological gel which exhibits huge contact-angle hysteresis.
The measured value of the surface-tension difference across the contact line is very similar to the liquid-vapor surface tension of the gelatin solution just before gelation.
This suggests that the surface tension of the interface between a gel and a droplet of its solvent can be negligible.

\begin{figure*}
    \centering
    \includegraphics[width=492pt]{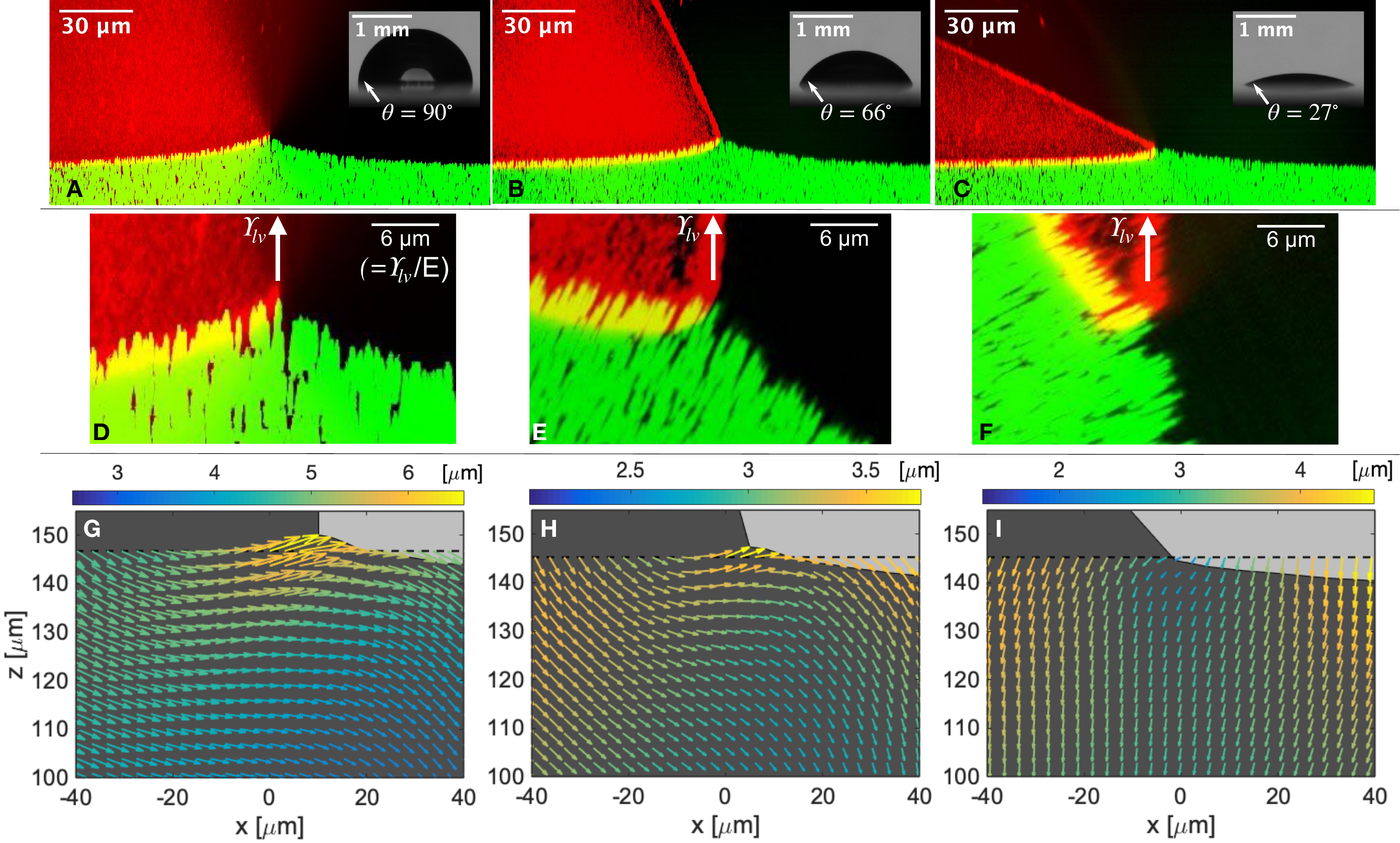}
    \caption{Sessile glycerol/water droplets on gelatin gels swollen with the same liquid. A droplet is deposited on gelatin, where its contact line pins so that liquid can be removed without moving the contact line.  (A-C) Side views of contact lines with different contact angles ($90^\circ, 66^\circ, 27^\circ$ respectively) imaged with macro photography (inset) and confocal microscopy,  where red/green nanoparticles label the droplets/gelatin, respectively. (D-F) shows a zoom in of the wetting ridges in (A-C), rotated so that  the liquid-vapor interface is vertical. 
    (G-I) shows the displacements under the ridge, calculated by tracking nanoparticle displacements between a relaxed state where the surface is flooded with liquid, and the deformed state under a contact line. The arrow colors indicate their magnitude.  }
    \label{fig:fluo_and_arrows}
\end{figure*}

For these experiments, we use a soft gelatin gel consisting of 5g of gelatin per 100ml of solvent phase, which is 4 parts glycerol to 1 part de-ionized water by volume.
This particular ratio was chosen, as it is essentially the same as a hydrogel, but shows minimal evaporation or hygroscopic behavior during experiments, avoiding the need for humidity control \cite{pare14}.
The gel has Young's modulus, $E=11.0 \pm 1.5$ kPa, Poisson ratio $\nu=0.39\pm 0.05$, and poroelastic diffusivity $D=3.0 \pm 1.5 \times 10^{-9}\mathrm{m^2}/\mathrm{s}$, as measured by indentation \cite{hu10}.
When we place a droplet of the solvent phase (surface tension $64~\mathrm{mN/m}$) on the surface of the cured gel, the droplet advances with a contact angle of $\theta\sim$130$^\circ$.
When liquid is subsequently removed with a pipette, the contact line remains pinned as $\theta\rightarrow 0$ (see insets to Figure \ref{fig:fluo_and_arrows}).
This enormous contact angle hysteresis allows us to create stable contact lines with a wide range of $\theta$.
In every experiment, we use fresh, never-wetted gelatin surfaces, so the surface outside the droplet's contact line has never been in contact with bulk liquid.

We ensure that there is sufficient time for the droplet to equilibrate with the substrate, and with the surrounding atmosphere by leaving it for $t_{eq}\approx 10^3~\mathrm{s}$ before imaging. 
This allows the gelatin to relax poroelastically over a region of size $\sqrt{Dt_{eq}} \approx 1~\mathrm{mm}$ from the contact line. This is much larger than the film thickness and the field of view under the contact line that we use during imaging, so there should be minimal residual poroelastic stresses to affect our results \cite{yoon2010poroelastic}.

The surface tension of the droplet strongly deforms the underlying material.
We place large, millimetric 4:1 glycerol:water droplets  on a thin film (147$\mu$m thick) of gelatin coated on a glass slide.
The films contain 0.003\% by volume of embedded, fluorescent, 100 nm-diameter nanoparticles. These do not segregate to the film surface, and are so dilute   that we do not expect their presence to affect the film's properties.
200 nm nanoparticles with a different emission spectrum are also attached to the glass slide, to act as reference points, and dispersed in the droplet. 
We image the sample from underneath with a 60x, 1.2 NA, water-immersion objective on a confocal microscope (further details in the Materials and Methods). 
This gives a 3-d intensity map of the substrate and the droplet.
We visualize the contact line by taking a maximum-intensity projection along a 30$\mu$m long, straight, pinning-defect-free section of contact line, which collapses the 3-D map into a side view of the contact line, as shown in Figure \ref{fig:fluo_and_arrows}A-C.
Here, particles in the droplet are shown as red, while the those in the substrate are green.
Note that the liquid is identical in both the droplet and gel phases.
The images clearly show the wetting ridge that is pulled up by the surface tension of the droplet, $\Upsilon_{lv}$.
As $\theta$ reduces from $\theta=90^\circ$ to $27^\circ$, the shape of the wetting ridge changes.

What can we extract from the shapes of the ridges? In previous work on hysteresis-free silicone gels, the angles between the three interfaces where they meet at the contact line were found to be fixed \cite{styl13}. 
This was interpreted as being due to a force balance between the surface tensions of the interfaces: $\Upsilon_{lv}, \Upsilon_{sv}$ and $\Upsilon_{sl}$. 
Then $\Upsilon_{sv}$ and $\Upsilon_{sl}$ could be extracted after measuring $\Upsilon_{lv}$ with standard techniques, and by applying the Neumann triangle construction \cite{styl13}.
This force balance is expected to hold only in a region around the contact line of size  $\ll\Upsilon_{lv}/E$, the elastocapillary length.
It is essential to distinguish surface energies, $\gamma$, from surface tensions (or surface stresses), $\Upsilon$.
The former represents the excess free energy of a material at its interface.
The latter are the measurable 2-dimensional stresses localized at the interface.
In simple liquids (such as the droplet here), $\gamma_{lv}=\Upsilon_{lv}$ \cite{styl17}.
However, for solids, $\gamma_s$ and $\Upsilon_s$ need not be the same, especially if $\gamma_s$ changes as the surface stretches \cite{styl17,xu18,andr20_review}.
In general, experiments that measure forces and displacements give access to surface tensions, as is the case here.

On gelatin, the microscopic geometry near the contact line appears to vary with $\theta$.
Figures \ref{fig:fluo_and_arrows}D-F show zoomed-in images of the tips of the wetting ridges from A-C, rotated so that the liquid-vapor interface is vertical.
The liquid-wedge angle seems to shrink dramatically as $\theta$ reduces.
If this represented changes in Neumann's triangle, this could mean that $\Upsilon_{sv}$ and $\Upsilon_{sl}$ are changing close to the contact line for different values of $\theta$.
On the other hand, the elastocapillary length, $\Upsilon_{lv}/E$, is only $6~\mathrm{\mu m}$ --
given the resolution of images in Fig \ref{fig:fluo_and_arrows}D-F, one should apply Neumann's triangle reluctantly.

Instead of focusing on the wetting ridge, we consider deformations away from the contact line, where strains are small and linear elasticity accurately captures the local balance of stress and strain.
To access the stress-free locations of the particles, we  flood the surface of the gel with the droplet phase at the end of the experiment. 
We track displacements of embedded nanoparticles as $\theta$ changes,
using a modified version of the tracker described in \cite{bolt17}.
Displacements are essentially zero in the direction parallel to the contact line.
Thus, we collapse the displacements along the contact line to give 2-d displacements maps, as shown in Figure \ref{fig:fluo_and_arrows}G-I.
There, the arrows start at (interpolated) nanoparticle positions ($x,z$) in the stress-free configuration, and end at the corresponding positions in the deformed state.
Displacements decay to zero at the rigid gelatin/glass interface at $z=0$, as shown in the full displacement fields in  the Supplement.

The displacement maps immediately reveal large differences under the contact line for the various $\theta$.
Most obviously, the magnitude of the displacements reduce significantly as the contact angle shrinks.
Surprisingly, for $\theta=90^\circ$ and $66^\circ$, despite the fact that the liquid-vapor interface pulls upwards, or to the left, there are strong displacements to the right underneath the contact line. For $\theta=27^\circ$, in-plane displacements reverse sign, and the contact line is pulled to the left. 
Interestingly, this would suggest that there is an intermediate angle where the average horizontal displacement vanishes -- potentially identifying a point where none of the horizontal component of $\Upsilon_{lv}$ is transmitted to the underlying substrate. 
Additionally, there is also a small overall shrinkage of the substrate between the deformed and reference states, as seen clearly in Figure \ref{fig:fluo_and_arrows}I.
There, all displacements point slightly downward, despite $\Upsilon_{lv}$ pulling up on the ridge.
This indicates that the flooding liquid slightly swells the film, suggesting  a slight  composition difference between the droplet and gel's solvent  (which had had time to equilibrate with the surrounding atmosphere).
Equivalently, there is a uniform gel pore pressure, $P$, that may vary between images.

\begin{figure}
    \centering
    \includegraphics[width=\columnwidth]{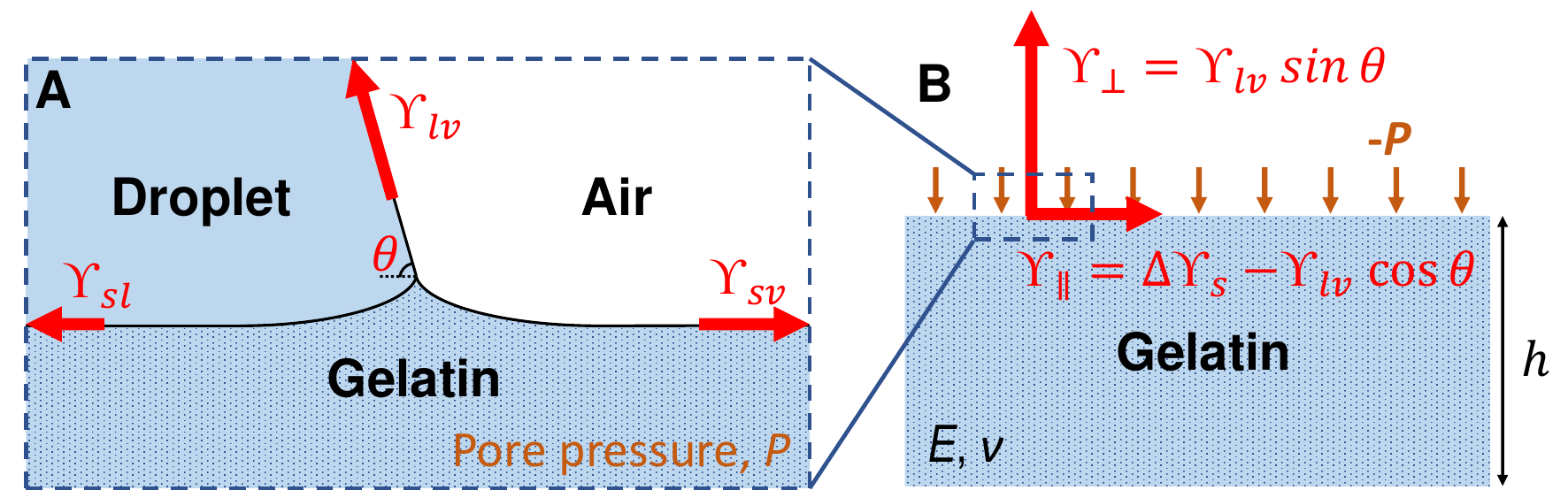}
    \caption{A schematic diagram showing the forces acting on the gelatin. (A) Near the contact line, there is a highly deformed wetting ridge. The gelatin may also be deformed by a pore pressure $P$, that acts to swell/deswell the gelatin (B) At a scale much bigger than that of the wetting ridge, the surface forces and pore pressure effectively act as horizontal and vertical line forces at the contact line, and as a uniform compressive pressure on the surface of a flat film.}
    \label{fig:schem}
\end{figure}

The displacement field gives us access to the stresses acting on the surface of the gel.
To characterize these, one option would be to convert displacements to strains, and then use the gel's poroelastic constitutive behavior to calculate the stress distribution in the substrate.
However, this requires a knowledge of $P$, which is not easy to measure.

Instead, we can exploit the symmetries of the problem to cleanly visualize the magnitude and orientation of the surface forces.
Figure \ref{fig:schem}A shows a schematic of the forces acting near the contact line.
Locally, there is a region where the surface is deformed, with a size of $\mathcal{O}(\Upsilon_{lv}/E=6~\mathrm{\mu m})$.
We zoom out to look from a much larger scale -- that of the  film-thickness, $h\gg \Upsilon_{lv}/E$, as shown in Figure \ref{fig:schem}B.
Provided the droplet's contact radius is $\gg h$, the substrate appears as a flat, linear elastic film, being acted on by vertical and horizontal linear forces at the contact line:
\begin{equation}
    \Upsilon_{\perp}=\Upsilon_{lv}\sin\theta
    \label{eqn:gammas1}
\end{equation}
and
\begin{equation}
 \Upsilon_{\parallel}=\Upsilon_{sv}-\Upsilon_{sl} - \Upsilon_{lv}\cos\theta \equiv \Delta\Upsilon  - \Upsilon_{lv}\cos\theta
    \label{eqn:gammas2}
\end{equation}
respectively, while the presence of a pore pressure is equivalent to a uniform vertical compression on the surface with a pressure $-P$.

To isolate the effects of $\Upsilon_\perp$ and $\Upsilon_\parallel$, we note that $P$  affects only vertical displacements of the substrate, $u_z(x,z)$, as it is uniform across the substrate surface, and acts directly downwards.
Thus, we focus only on the horizontal displacements, $u_x(x,z)$.
Furthermore, we notice that $\Upsilon_\perp$ results in horizontal displacements that are symmetric about the position of the contact line ($x=0$),
while  $\Upsilon_\parallel$ results in anti-symmetric horizontal displacements about $x=0$ 
(see the Appendix for a proof).
Thus, if we decompose $u_x$ into odd and even parts $u_x^{o}=(u_x(x,z)-u_x(-x,z))/2$, and $u_x^{e}=(u_x(x,z)+u_x(-x,z))/2$ respectively, the resulting displacements take the form
\begin{align}
\label{eqn:us}
    u_x^{o} &=\frac{\Upsilon_\perp}{E}  U_\perp(x,z,\nu) \\
    \,u_x^{e} &=\frac{\Upsilon_\parallel}{E} U_\parallel(x,z,\nu),
\end{align}
where the functions $U_\perp$ and $U_\parallel$ are given in the Supplement (following \cite{xu10}). 
Thus, if the experimentally measured $u_x^{o}$ and $u_x^{e}$ match the patterns of $U_\perp$ and $U_\parallel$, we can use their relative magnitudes to extract $\Upsilon_\perp$ and $\Upsilon_\parallel$.
Note, however, that we expect the patterns to deviate in a region around the contact line of size $\mathcal{O}(\Upsilon_{lv}/E)$), due to contributions from solid surface tension \cite{styl17}.

\begin{figure*}
    \centering
    \includegraphics[width=500pt]{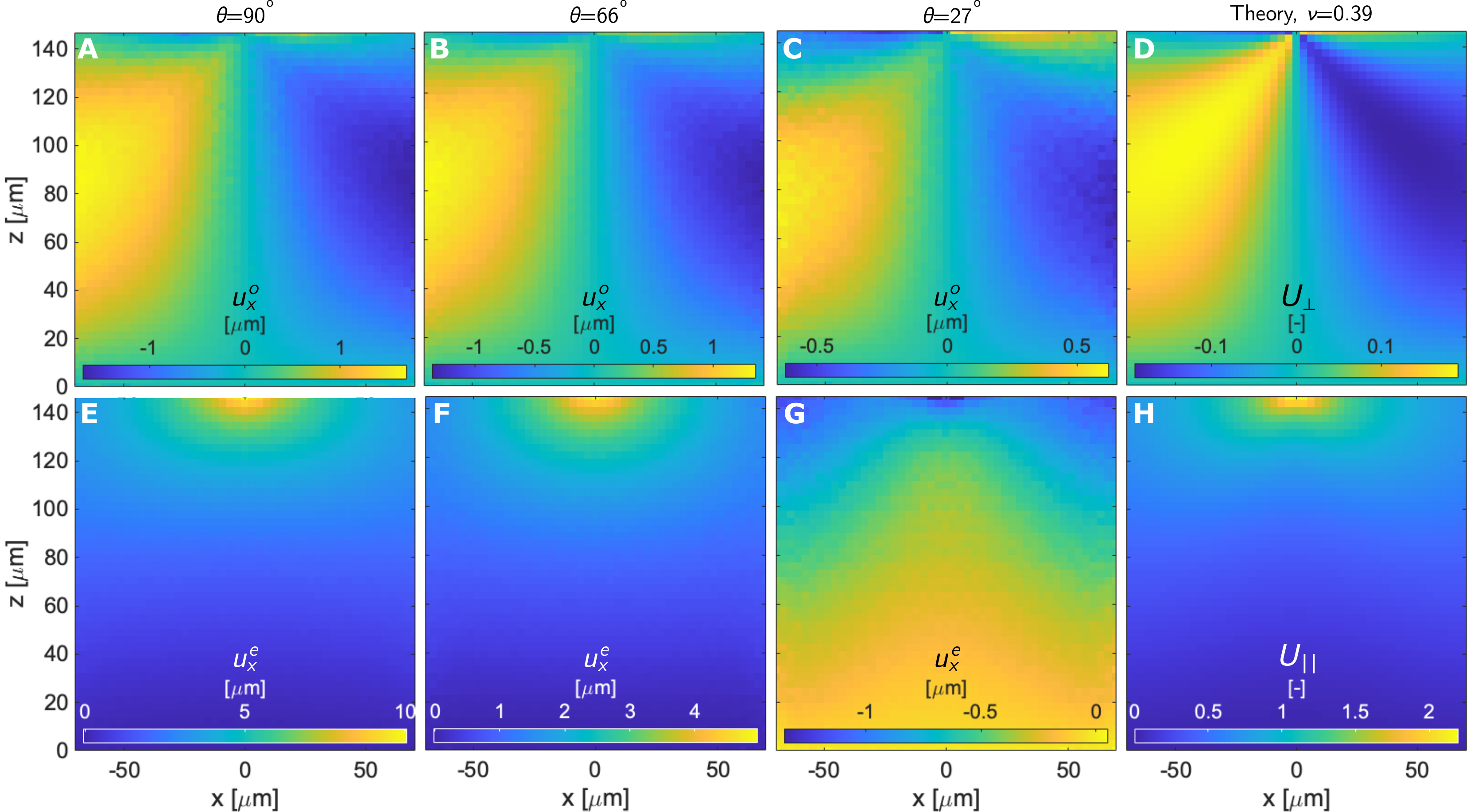}
    \caption{Decomposing the horizontal displacements under a contact line. The top/bottom rows show the odd/even parts of $u_x(x,z)$: $u_x^o$ and $u_x^e$. The first three columns (A-C,E-G) correspond to $\theta=90^\circ,66^\circ,27^\circ$ respectively. The final column (D,H) shows $U_\perp$ and $U_\parallel$ from equations (\ref{eqn:us}), using $\nu=0.39$ and $h=147\mu$m.
    The experiments show good agreement with the theory. Note that the magnitude of (G) is negative, causing the colors to be the reverse of those in (E,F,H).
    }
    \label{fig:sym_asym}
\end{figure*}

Indeed, the decomposed displacement fields match well with linear-elastic predictions for $U_{\parallel,\perp}$.
Figure \ref{fig:sym_asym} shows $u_x^{o}$ (A-C) and $u_x^{e}$ (E-G) for the different $\theta$ in Figure \ref{fig:fluo_and_arrows}, while (D,H) show $U_\perp$ and $U_\parallel$.
These show strong qualitative agreement, including  $u_x^e$ for $\theta=27^\circ$, where the sign of the displacements is reversed. 
\begin{figure}
    \centering
    \includegraphics[width=\columnwidth]{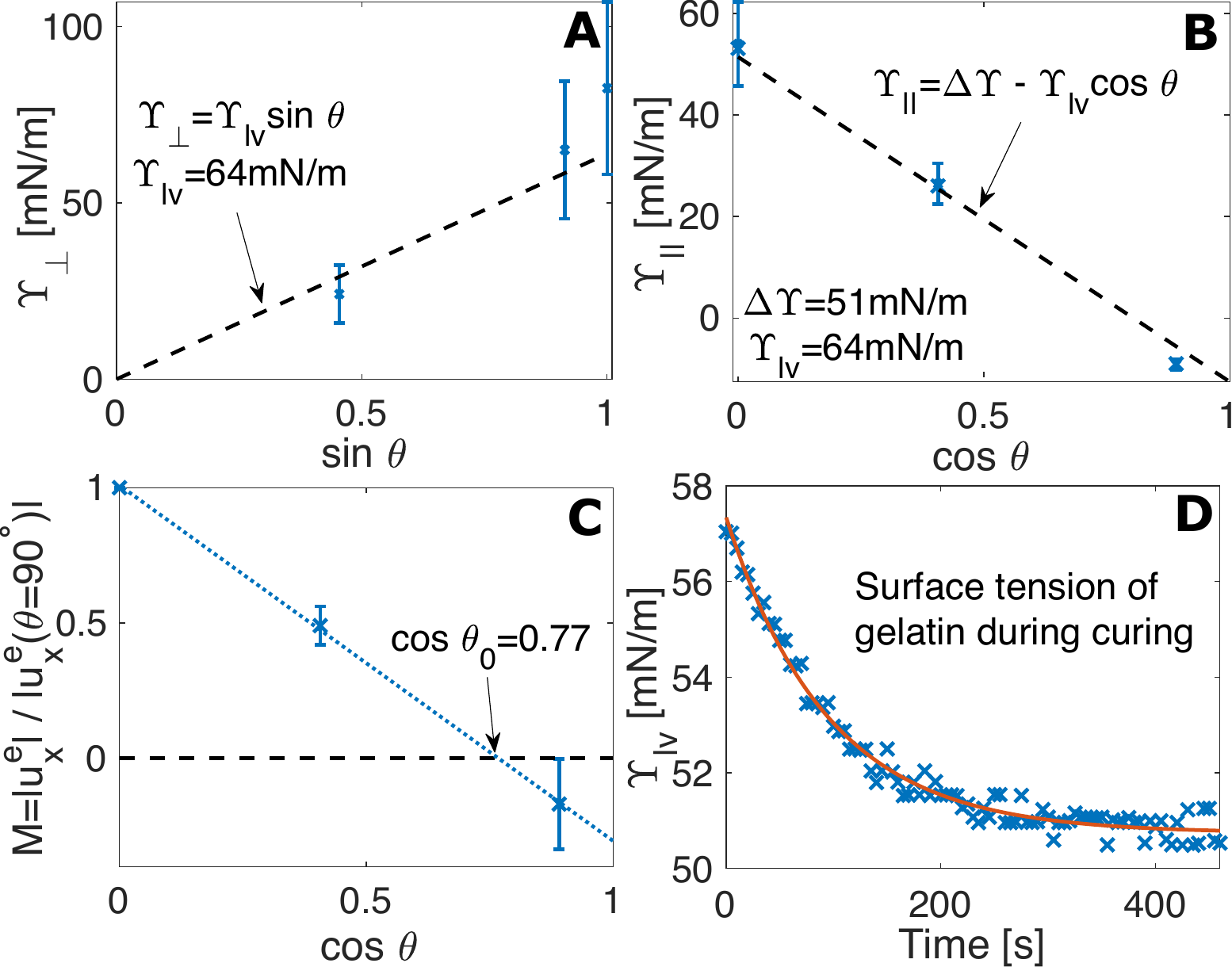}
    \caption{Measuring surface tensions on the gel. (A,B) show $\Upsilon_\perp(\theta)$ and $\Upsilon_\parallel(\theta)$, measured by fitting displacement fields to the theoretical expressions in (\ref{eqn:us}). The dashed line shows the theoretical predictions from (\ref{eqn:gammas1},\ref{eqn:gammas2}).
    (C) Shows $M$, the relative magnitudes of the displacement fields in Figure \ref{fig:sym_asym}E-G. The angle, $\theta_0$, where $M=0$ corresponds to the point where the stresses acting at the gel's surface balance horizontally. 
    (D) The surface tension of a curing gelatin droplet, measured \emph{via} pendant-droplet tensiometry. The red curve shows an exponential fit to the data, which plateaus to a value of 51 mN/m.}
    \label{fig:extracted_upsilon}
\end{figure}

We extract $\Upsilon_\perp$ and $\Upsilon_\parallel$ using a least-squares fitting of the odd and even displacement fields to the theoretical expressions (Eq. \ref{eqn:us}).
We use the entire image for fitting, as this gives essentially the same results as when we ignore the region close to the contact line (\emph{e.g.} within a distance $3\Upsilon_{lv}/E$).
Images showing residual errors between the data and the theory are given in the Supplement for each data set.
The resulting values of $\Upsilon_\perp$ and $\Upsilon_\parallel$
are shown in Figure \ref{fig:extracted_upsilon}A,B.
The errorbars reflect the measured uncertainty in $E$ and $\nu$.
Note that these are much larger for $\Upsilon_\perp$ than $\Upsilon_\parallel$.
Measurement of $\Upsilon_\perp$ is much more sensitive to the value of $\nu$ (which has relatively wide errorbars), as it involves substrate contractions perpendicular to the loading direction.

Furthermore, the fitted values of $\Upsilon_\perp$ and $\Upsilon_\parallel$ agree well with the expected forms given in Eqs (\ref{eqn:gammas1},\ref{eqn:gammas2}). 
Using the value of $\Upsilon_{lv}=64$mN/m, measured by hanging drop tensiometry, we plot these equations as dashed lines in Figure \ref{fig:extracted_upsilon}A,B.
For $\Upsilon_\perp$, we see agreement with the data with no fitting parameters.
For $\Upsilon_\parallel$, there is one fitting parameter, $\Delta\Upsilon$, the difference in the solid-vapor and solid-liquid surface tensions.
For these data, we find $\Delta \Upsilon=51 \pm 3~\mathrm{mN/m}$.
Note that $\Delta \Upsilon$ could also be extracted with data from a single contact angle, $\theta$, using this approach.

When the linear elastic properties of the gel ($E$ and $\nu$) are unknown, a series of measurements with different contact angles can still reveal $\Delta\Upsilon$.
The key is the dependence of the magnitude of $u_x^{e}$ on the contact angle  $\theta$. 
At some angle, $\theta_0$, $u_x^{e} = 0$. 
At this point, $\Upsilon_\parallel=0$, and  $\Delta\Upsilon=\Upsilon_{lv}\cos\theta_0$, as given by Eq. \ref{eqn:gammas2}. 
We determine $\theta_0$ by plotting the relative magnitude, $M$, 
of $u_x^{e}(\theta)$ and $u_x^{e}(\theta=90^\circ)$ against $\cos\theta$ in Figure \ref{fig:extracted_upsilon}C.
$M$ is defined as the constant that minimizes $(u_x^{e}(\theta=90^\circ)-Mu_x^{e}(\theta))^2$ over all the pixels in Figure \ref{fig:sym_asym}.
Then we fit a line $M=1-\alpha \cos\theta$ to the data, and extract $\cos\theta_0=1/\alpha=0.77$.
This yields $\Delta\Upsilon= 49\pm 5~\mathrm{mN/m}$, which is in  good agreement with the values that we extracted  from the mechanical model using the elastic properties of the gel. 

To corroborate these results, we measured the surface tension of gelatin solutions as they cured using pendant-drop tensiometry \cite{del97}.
Gelatin, dissolved at high temperature, does not instantly cure upon cooling to room temperature. Instead it takes $\mathcal{O}(20\mathrm{mins})$ to gel \cite{rons17}.
Thus, we can hang a liquid droplet of curing gelatin at room temperature, and measure its surface tension as it cross-links, as shown in Figure \ref{fig:extracted_upsilon}D.
Note that this method, which assumes hydrostatic equilibrium, is only applicable as long as the gelatin remains fluid.  Therefore, it can reliably report surface tension on the approach to the gel point, but not past it. 
As gelation proceeds, the surface tension reduces.  
After around 5 minutes, the apparent surface tension plateaus to a final value of 51$\pm$1 mN/m.
With the assumption that the surface is not significantly altered across the gel point, this suggests that $\Upsilon_{sv} \approx51$ mN/m.
In the gel state, we further anticipate that 
 $\Upsilon_{sl} \ll \Upsilon_{sv}$, as both sides of the liquid/gel interface are predominantly the same solvent, and the concentration of the polymer is very dilute, only 3-4\%. 
Together, these suggest that $\Delta\Upsilon=51\pm1~\mathrm{mN/m}$, in excellent agreement with our gel deformation measurements.

At the angle $\theta_0$, the liquid vapor interface transmits no in-plane forces to the substrate, $\Upsilon_\parallel=0$.  According to Eq. \ref{eqn:gammas2}, this mechanical equilibrium is given by
\begin{equation}
    \Upsilon_{lv}\cos{\theta_o} =\Upsilon_{sv}-\Upsilon_{sl}.
    \label{eq:yd2}
\end{equation}
It is independent of the bulk mechanical properties, such as $E$ and $\nu$, and the far field boundary conditions such as the film thickness, $h$. 
This is similar in form to the law of Young-Dupr\'{e}, Eq. \ref{eq:young}.
While $\theta_{eq}$ reveals  surface energies,  $\theta_0$ reveals  surface tensions localized to the contact line.
These two are equivalent ($\theta_0=\theta_{eq}$)  when $\Upsilon_{sv}-\Upsilon_{sl}=\gamma_{sv}-\gamma_{sl}$.
This is the typical assumption for gels, since they are mostly made of a simple liquids   \cite{mora10,chak13,schu18,liang2018surface,masu19}.
However, any solid material can, in principle, have different values of  $\gamma $ and $\Upsilon$ \cite{shut50}.
The relevance of this distinction for swollen polymer networks remains an actively debated topic.

\begin{figure}
    \centering
    \includegraphics[width=\columnwidth]{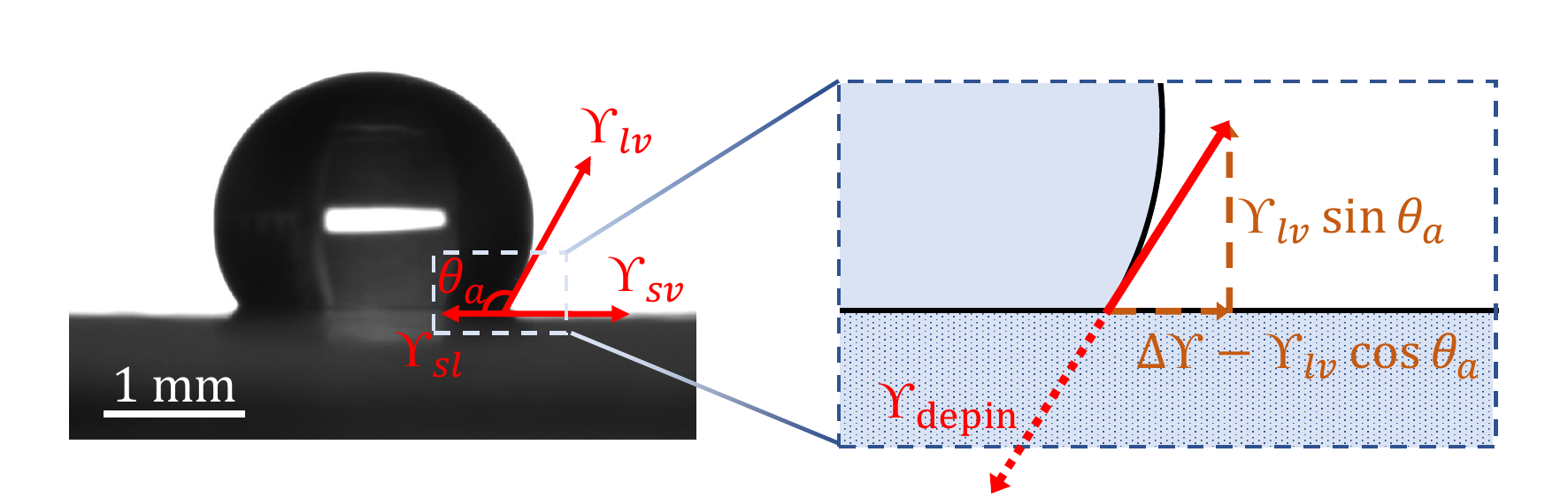}
    \caption{The depinning force at the contact line at the advancing contact angle. This opposes the surface-tension forces from $\Upsilon_{lv}$, $\Upsilon_{sv}$ and $\Upsilon_{sl}$. Note that, in general, this is not parallel to the liquid surface at the contact line.}
    \label{fig:depin}
\end{figure}

Our results allow us to calculate the maximum force that can be exerted on the substrate before depinning occurs -- something that can not be measured without an accurate value of $\Delta\Upsilon$.
Consider a droplet at its advancing contact angle, $\theta_a$, as shown in Figure \ref{fig:depin}.  
Evaluating  Eqs (\ref{eqn:gammas1},\ref{eqn:gammas2}) in this configuration yields the total magnitude of the depinning force
\begin{equation}
    \Upsilon_{\mathrm{depin}}=\sqrt{\Upsilon_{lv}^2+(\Delta\Upsilon)^2-2\Upsilon_{lv}\Delta\Upsilon \cos\theta_a}.
\end{equation}
For the current experiments, this gives  $\Upsilon_\mathrm{depin}=104~\mathrm{mN/m}$.
It is interesting to speculate what sets this value, as hysteresis can arise for multiple reasons, including surface roughness, microscopic chemical heterogeneity of the surface, molecular reorientation at the surface, and dynamic hysteresis due to viscoelastic or poroelastic braking \cite{erbi99}.
Here, it is unlikely that there is significant surface roughness of the native gel surface, as surface tension keeps it extremely flat during the curing process.
Viscoelastic braking \cite{carr96,lhermerout2016moving} or poroelastic braking \cite{guan2020state} is also unlikely here, as droplet contact angles show little relaxation over the course of days.
Instead, we expect it to be related to processes at the surface including yielding and damage \cite{baum06,grze17}, molecular reorientation of the surface \cite{holl75,holl76,yasu81,butt18}, or potentially strain-dependent surface properties \cite{snoe18,van18}.
Fully understanding the origins of pinning here is beyond the scope of this paper. 
However, we believe our approach provides important new tools to explore this question. 

In summary, we present a new approach for the characterization of surface properties by wetting experiments.  Traditionally, the contact angle of the droplet is interpreted to reveal interfacial free energy differences. While this is on the firmest conceptual foundations, practically the equilibrium contact angle can only be narrowed down within a range set by the advancing and receding contact angles.  On all but ideal surfaces, this range is too broad to pin down  surface properties.  We argue for a different approach, based on the less stringent requirement of mechanical, rather than thermodynamic, equilibrium.  Specifically, we have shown that it is possible to determine the contact angle where the net in-plane forces on the substrate is zero. In this way, we can quantify differences in interfacial tension, even in the presence of large contact-angle hysteresis. This generalizes the concept of the equilibrium contact angle, and is captured in compact form by Eq. \ref{eq:yd2}, which echoes the original formulation of Young and Dupr\'{e}.

Our approach enables the measurement of surface-tension differences and depinning forces at contact lines on soft solids, even in the presence of huge contact-angle hysteresis. Even when the solid's elastic properties are unknown.
As we expect that the surface tension between a gel and its own solvent ($\Upsilon_{sl}$) is minimal, this effectively allows us to evaluate absolute values of $\Upsilon_{sv}$.
Our approach can be used on relatively stiff substrates: potentially up to  $\mathcal{O}(100\mathrm{kPa})$, when using light microscopy to track displacements.
This is in contrast to elastocapillary approaches limited  to softer materials (\emph{e.g.} \cite{styl13,chak13,mora10,mora11,shao20}).
Additionally, our technique does not restrict one to immiscible substrate/liquid pairings (\emph{c.f.} \cite{styl13,nade13,mond15}), and the measured surface stress values should be for the fully-relaxed state, unlike other options, which extract surface stress values in situations where the surface is significantly deformed \cite{mora10,chen12,styl13,xu17}.
Although we have used the entire displacement field under a contact line, similar information should be measurable  from the displacements at the surface away from the contact line.
Further, here, we only consider the common case where a droplet is much larger than the substrate layer thickness, but we note that smaller probe droplets can certainly be used by adapting the theory presented above to use previous analytic solutions \cite{styl12}.

We anticipate that future work can extract further information about the system from the near-contact-line region by comparison with analytic models that account for different types of behavior at the gel surface \cite{bost14,bard18,styl18mecheqbm}.
Our technique could also be used to measure mechanical substrate properties with one single measurement (\emph{c.f.} \cite{gros17} for alternative approaches), as vertical displacements of the substrate contain information about the pore pressure in the film, while the detailed pattern of substrate deformations depends on $\nu$ and $E$.
Importantly, it is well-known that surface properties of soft materials can be history-dependent.
Thus, we anticipate that measured value of surface tensions, and the contact angle representing mechanical equilibrium, may change for advancing and receding contact lines.
For such studies, systematic variation of liquid-substrate combinations will be essential.

\begin{acknowledgments}
R.W.S. is supported by the Swiss National Science foundation (grant 200021-172827). J.Y.K. is supported by the MOTIE in Korea, under the Fostering Global Talents for Innovative Growth Program supervised by the KIAT (grant P0008746). We acknowledge helpful conversations with Katrina-Smith Mannschott  and Anand Jagota.

\end{acknowledgments}

\appendix

\section{Materials and Methods}

To make the gelatin gels, we dissolved 5g gelatin (General purpose grade gelatine, Fisher) in a mixture of 80 mls glycerol (VWR) and 20 mls de-ionized water. The mixture was heated with stirring at 90$^{\circ}$ for 30 minutes (following \cite{baum06}).
After heating, 100$\mu$l of 100nm-diameter, carboxylate-modified, red fluorescent nanoparticles  (2\% solids, Fluospheres, ThermoFisher Scientific) were added into the mixture, and it was poured into moulds and allowed to cool for 24 hours at 4$^{\circ}$C before being used.
For measuring mechanical properties, we filled 1cm-deep petri dishes with the gelatin.

For confocal imaging, we created thin films of gelatin on 50mm petri dishes with a No. 1.5 glass cover slip as their bottom surface (P50G-1.5-30-F, MatTek).
This involved first attaching yellow-green fluorescent nanoparticles to the glass surface, which act as fixed reference points for tracking the bottom of the gelatin layer, and then adding the gelatin layer.
To attach the reference particles to the glass surface, we follow the protocol described in \cite{styl14}.
In brief, we activate the surface in a UV-ozone cleaner, functionalize it via vapor deposition of  (3-Aminopropyl)triethoxysilane, and then submerge it in a solution containing 200nm-diameter, yellow-green, carboxylate-modified, fluorescent nanoparticles (2\% solids, Fluospheres, ThermoFisher).
We then rinse this with de-ionized water.
To make the thin films of gelatin, we placed 30$\mu$l of the hot gelatin solution on the bead-coated glass, and then covered this with a 18mm-diameter, \#1.5 circular cover glass (Paul Marienfeld, GmbH).
After curing, the cover glass was gently removed with tweezers, and the substrate was used for experiments.

For imaging of contact lines, we make a solution containing 8ml glycerol, 2ml de-ionized water and 10ul of the (2\% solids) yellow-green nanoparticles. 
Then we place a $5\mu$l droplet of the solution on the gelatin layer.
We reduced the size of the droplet when required by manually removing liquid with a pipette, approximately 1$\mu$l at a time.
After each change in droplet volume, we let it rest for at least 20 minutes before imaging.

3D confocal imaging was done on a microscope (Nikon, Eclipse Ti2) with a spinning disk confocal system (Yokagawa CSU-X1) using a 60x water immersion objective lens (Nikon, MRD07602). We used 488 nm and 560 nm lasers for yellow-green and red fluorescent nanoparticles respectively.
Images were acquired with at 330nm intervals in $z$ to obtain 3-d measurements. 

We measure the surface tension of liquids using a homemade pendant-droplet tensiometry setup.
Droplet shapes are analysed with axisymmetric drop shape analysis \cite{del97}.
For the droplet phase, we measure $\Upsilon_{lv}$ of the pure liquid, and liquid that has been allowed to equilibrate on the gelatin substrate for an hour.
This gives $\Upsilon_{lv}=66\pm 1 $mN/m and $64\pm 1$mN/m respectively, suggesting that there is very little contamination of the droplet by material from the gelatin.

\section{Odd/even responses to line loadings}
For a vertical line force, $\Upsilon_{\perp}$, acting on the substrate as shown in Figure \ref{fig:schem}B, there is complete left/right symmetry to the problem.
This immediately implies symmetric inward/outward displacements, so that the resulting displacement field, $u_x^\perp(x,z)$, is odd.

For a horizontal line force, $\Upsilon_\parallel$, let the solution be $u_x^\parallel(x,z,\Upsilon_\parallel)$.
The solution for $\Upsilon_\parallel\rightarrow - \Upsilon_\parallel$ satisfies $u_x^\parallel(x,z,\Upsilon_\parallel)+u_x^\parallel(x,z,-\Upsilon_\parallel)=0,$
as the sum of the line forces $\Upsilon_\parallel$ and $ - \Upsilon_\parallel$ are zero, so the sum of the resulting displacement fields must also be zero.
Furthermore, the solution for a negative line force is the same as the solution for a positive line force, reflected about $x=0$. Mathematically this can be expressed as $u_x^\parallel(x,z,-\Upsilon_\parallel)=-u_x^\parallel(-x,z,\Upsilon_\parallel)$.
Combining these two equations gives $u_x^\parallel(x,z,\Upsilon_\parallel)=u_x^\parallel(-x,z,\Upsilon_\parallel)$, so $u_x^\parallel$ is even.



%

\end{document}